\documentclass[reprint,
               superscriptaddress,
               floatfix,
               amsmath,
               amssymb]{revtex4-2}

\usepackage{graphicx}
\usepackage{float}
\usepackage{svg}
\usepackage[mode=tex]{standalone}
\usepackage{tikz}
\usetikzlibrary{shapes}
\usetikzlibrary{positioning}
\usetikzlibrary{calc}
\usetikzlibrary{fit}
\usetikzlibrary{arrows.meta, shapes.misc}

\usepackage{xparse}

\definecolor{c0}{HTML}{0173B2}
\definecolor{c1}{HTML}{029E73}
\definecolor{c2}{HTML}{D55E00}
\definecolor{c3}{HTML}{CC78BC}
\definecolor{c4}{HTML}{ECE133}
\definecolor{c5}{HTML}{56B4E9}
\definecolor{c6}{HTML}{e377c2}
\definecolor{c7}{HTML}{7f7f7f}
\definecolor{c8}{HTML}{bcbd22}
\definecolor{c9}{HTML}{17becf}

\tikzset{
dot/.style = {circle, fill=black, minimum size=#1,
              inner sep=0pt, outer sep=0pt},
dot/.default = 6pt 
}

\usepackage{dcolumn}
\usepackage{bm}
\usepackage{bbold}

\usepackage{pifont}
\newcommand{\cmark}{\ding{51}}%
\newcommand{\xmark}{\ding{55}}%

\definecolor{linkred}{RGB}{160,0,0}
\definecolor{citegreen}{RGB}{0,160,0}
\definecolor{urlblue}{RGB}{0,0,160}
\usepackage[pdfstartview=FitH,      
            breaklinks=true,        
            bookmarksopen=true,     
            bookmarksnumbered=true, 
            colorlinks=true, urlcolor=urlblue, pdfborder={0 0 0},
            linkcolor=linkred,
            citecolor=citegreen,
            pdfencoding=auto,
            psdextra
            ]{hyperref}

\usepackage{braket}
\let\vec\bm
\usepackage[obeyFinal,section]{easy-todo}
\usepackage{cleveref}

\usepackage[ruled,vlined]{algorithm2e}
\SetKwComment{Comment}{$\triangleright$\ }{}

\SetCommentSty{mycommfont}

\usepackage{pgfplots}
\graphicspath{{figures/}{.}}
\makeatletter
\def\input@path{{figures/}{.}}
\makeatother

\pgfplotsset{compat=1.16}
\providecommand\inputpgf[2]{{
\let\pgfimageWithoutPath\pgfimage
\renewcommand{\pgfimage}[2][]{\pgfimageWithoutPath[##1]{#1/##2}}
\input{#1/#2}
}}

\newcounter{jlbcommentcounter}

\newcounter{amcommentcounter}

\newcounter{rscommentcounter}

\newcommand{\EVQE}{E_{\textnormal{VQE}}}

\newcommand*\diff{\mathop{}\!\mathrm{d}}

\newcommand{\dk}{\frac{\diff{\vec{k}}}{(2\pi)^d}}

\begin{document}

\preprint{APS/123-QED}

\title{Sketching phase diagrams using low-depth variational quantum algorithms}

\author{Jan Lukas Bosse}
  \email{janlukas.bosse@bristol.ac.uk}
\affiliation{School of Mathematics, University of Bristol}
\affiliation{Phasecraft Ltd.}
\author{Raul A. Santos}%
 \email{raul@phasecraft.io}
\affiliation{Phasecraft Ltd.}
\author{Ashley Montanaro}%
 \email{ashley@phasecraft.io}
\affiliation{Phasecraft Ltd.}
\affiliation{School of Mathematics, University of Bristol}

\date{\today}

\begin{abstract}
  Mapping out phase diagrams of quantum systems using classical simulations can be 
  challenging or intractable due to the computational resources required to simulate even 
  small quantum systems far away from the thermodynamic limit. We investigate 
  using quantum computers and the Variational Quantum Eigensolver (VQE) for this 
  task. 
  In contrast to the task of preparing the exact ground state using VQE, sketching 
  phase diagrams might require less quantum resources and accuracy, 
  because low fidelity approximations to the ground state may be enough to 
  correctly identify different phases.
  We used classical numerical simulations of low-depth VQE circuits to compute
  order parameters for four well-studied spin and fermion models which
  represent a mix of 1D and 2D, and exactly-solvable and classically hard
  systems.
  We find that it is possible to predict the location of phase transitions up
  to reasonable accuracy using states produced by VQE even when their overlap
  with the true ground state is small.
  Further, we introduce a model-agnostic predictor of phase transitions based
  on the speed with which the VQE energy improves with respect to the circuit
  depth, and find that in some cases this is also able to predict phase
  transitions.
\end{abstract}

\maketitle


\section{Introduction} 
\label{sec:introduction}

In a quantum system composed of a large aggregate of particles, a complete
description of the full state vector is impractical. A useful approach at the
basis of statistical mechanics consists of describing the system in terms of a
few properties that describe the collective behaviour of the particles of the
system. Abrupt changes in these descriptors as a function of external
parameters signal transitions between different phases.
At zero temperature, transitions driven by changes of some set of parameters
$\vec{g}$ in a Hamiltonian $H(\vec{g})$ are known as quantum phase transitions
and are fully determined by the ground state of the system
$\ket{\psi_0(\vec{g})}$. The relevant descriptors are the ground state 
expectation values of some (local) observables $\hat{O}_i$.
Thus we are interested in characterising the
expectation values of a set of observables $\hat{O}_i$ in this ground state, i.e.
$O_i(\vec{g})=\braket{\psi_0(\vec{g})|\hat{O}_i|\psi_0(\vec{g})}$

In computational solid state and condensed matter physics
zero-temperature phase diagrams are usually obtained by finding the ground
state of the system, using e.g.\ exact diagonalisation, Monte Carlo methods, or
tensor network methods like DMRG or variational methods for PEPS and MERA
ansatz states and then measuring the order parameters $O_i(\vec{g})$ in that ground state. The
behaviour of these order parameters then gives information about the different
phases the ground state may lie in~\cite{deJongh_1998,Fradkin_1978,Laeuchli_2006,Buchta_2005}. 

Quantum computers could allow the production of families of approximate ground
states that are not accessible classically. One method for producing such
states, which is particularly well-suited to the noisy intermediate-scale
quantum (NISQ) regime, is the Variational Quantum Eigensolver
(VQE)~\cite{Peruzzo2014}. VQE is a hybrid quantum-classical algorithm to
produce a ground state of a quantum Hamiltonian $H$ via the variational
principle. A classical optimiser is used to minimise the expectation value
$\braket{\psi(\vec\theta) |H| \psi(\vec\theta)}$ over a family of states
$\ket{\psi(\vec\theta)}$. The hope is that this family is sufficiently
expressive that there is some choice of $\vec\theta$ such that
$\ket{\psi(\vec\theta)}$ is an approximate ground state of $H$.

A natural method for applying VQE to approximately computing phase diagrams is
then to attempt to find the ground state of $H(\vec g)$ at different parameters
$\vec g$ using VQE and then measure the order parameter on the quantum computer
in that state. If VQE has produced a high-fidelity approximation to the ground
state, the measured value will be close to the true order parameter value.
However, even if the approximation produced by VQE is low-fidelity, it is still
possible that the measured order parameter will be accurate. This is because
order parameters are often local observables that may have the same expectation
value in two states, even if the fidelity between those states is low and their
global structure is different.

One constraint to this approach is that it requires prior knowledge of the order
parameter to be measured, which depends on some knowledge of the physics of the
system. We therefore also consider a different kind of order parameter: the
rate at which the VQE energy improves as the circuit depth $p$ is increased.
Compared to measuring order parameters in the VQE state, this method has the
advantage of being model agnostic, as one does not need to understand the
physics of the model in question to use the correct order parameters, and the
only data required are the VQE energies for different circuit depths throughout
the phase diagram. We also test the limitations of this procedure in cases where
accessing local information cannot distinguish different phases, like in the
case of topological phase transitions.

This approach is in spirit similar to the work of Mondaini et al.~\cite{Mondaini_2022}
who use the average sign in quantum Monte-Carlo methods as a model agnostic 
indicator of different phases.
We further motivate our approach by two intuitions. First, ground states of
quantum systems at a phase transition are characterised by long-range
entanglement~\cite{Sachdev_2011}, and creating complex entangled states will in general
require high-depth VQE circuits. Second, Chen, Gu and Wen~\cite{Chen_2010} have
shown that symmetry-preserving, local low-depth circuits can map ground states
of local Hamiltonians to ground states of other local Hamiltonians, if and only
if there is a symmetry-preserving adiabatic path connecting the two Hamiltonians.
In \cref{sec:vqe_hardness_in_noninteracting_free_fermion_systems}, we strengthen 
this result for translation-invariant non-interacting fermionic models and show
that deep VQE circuits with the Hamiltonian Variational
ansatz~\cite{Wecker2015} are required to prepare ground states at or near
criticality. We also show that the exact circuit depth for ground states near
criticality depends on the gap of the Hamiltonian and hence deeper circuits are
required closer to the critical point.

Further support for this approach is provided by recent work of Dreyer, Bejan
and Granet~\cite{Dreyer_2021}, who analytically find the optimal VQE parameters
and energies for the 1D transverse field Ising model with the Hamiltonian
variational ansatz~\cite{Wecker2015}, and show that when the initial state for VQE and the target
Hamiltonian are in the ordered phase, the VQE energy error (the difference
between the true ground state energy and the lowest energy found by VQE) scales
with the ansatz depth $p$ as $e^{-\lambda p}$ whereas it scales with $p^{-1}$
if the target Hamiltonian is in the disordered phase and with $p^{-2}$ at the
critical point. In more recent, related work Roca-Jerat et al.~\cite{RocaJerat_2023} 
and Jayarama and Svensson~\cite{Jayarama_2022}
also study the complexity of different state preparation methods, including 
variational algorithms, and find that the complexity depends crucially on whether 
or not the system passes close to a quantum critical point when preparing the 
state. Related to our work is also that of Okada et. al.~\cite{Okada_2022} who 
propose optimising low-depth VQE circuits on a classical computer and then using 
the quantum computer only to measure non-local order parameters to 
identify topological phases.

\textbf{Our results.}
After briefly reviewing the Variational Quantum Eigensolver method 
in~\cref{sec:the-variational-method} and various methods for warm starting 
optimisation from previous VQE runs in~\cref{sec:warm_starting_optimisation_with_previous_results}
we present our numerical results in~\cref{sec:results}.
We compare the approach of measuring order parameters in the VQE state and the
approach of analysing the VQE energy derivative to detect and locate phase
transitions using VQE using a number of well-understood models. First
in~\cref{sec:the_tfim} is a spin-1/2 chain, the transverse field Ising model
which has a simple and completely understood phase diagram. Secondly
in~\cref{sec:the_bb_chain}, we considered a spin-1 chain, the
bilinear-biquadratic model~\cite{Laeuchli_2006,Buchta_2005}, with a richer but
still well-understood phase diagram. Finally in~\cref{sec:the_2d_ssh_model}, we
also studied a free fermion model in two dimensions, the 2D SSH model as
studied in~\cite{Obana_2019}. This model is exactly solvable, enabling us to
carry out simulations on a $10\times 10$ lattice (100 qubits).
Additional data for all three models can also be found in~\cref{sec:more_data}.

\begin{table}
  \begin{center}
    \begin{tabular}[c]{l|c|c}
      \multicolumn{1}{c|}{\textbf{}} & 
      \multicolumn{1}{c|}{\textbf{Order parameters}} &
      \multicolumn{1}{c}{\textbf{VQE energy derivative}} \\
      \hline
      1D TFIM & \cmark & \cmark \\
      2D TFIM & \cmark & \cmark\\
      1D BBC & \cmark & \cmark / \xmark\\
      2D SSH & \cmark & \xmark \\
    \end{tabular}
  \end{center}
  \caption{Comparison of the performance of the different methods used to 
  locate phase transitions across the different models studied by us.
  A check mark means that the corresponding method signalled phase transitions 
  via an extremum, discontinuity or discontinous first derivative.
  Because in the case of the bilinear-biquadratic chain the VQE energy 
  derivative detected some, but not all phase transitions we put a check mark 
  and cross on the corresponding entry of the table. Due to 
  finite size effects it is hard to further quantify the success.
  }
  \label{tab:result_overview}
\end{table}

We find that both methods -- searching for the ground state using VQE and then
measuring order parameters, and analysing the VQE energy derivative -- are
viable ways to identify phase transitions in the systems we studied. However,
neither method is effective for all phase transitions in all systems we
studied. For the transverse field Ising model, both methods correctly predict
the phase transition and the VQE energy derivative seems to be less prone to
finite-size errors than the magnetisation. For the bilinear-biquadratic chain,
the VQE energy derivative correctly locates three of the five phase
transitions while the order parameters corresponding to the different phases
correctly detect those, but have marked finite-size errors in the small system
sizes that we simulated. For the 2D SSH model the behaviour is less conclusive
and the VQE energy derivative fails to give a clear signal at the phase
transition, while an appropriately chosen order parameter measured in the VQE
state correctly identifies the phase transition. This indicates that the
VQE energy derivative cannot capture topological phase transitions.
Besides the VQE energy derivative we also experimented with fitting different
functions to the curve $\EVQE(p)$ to detect phase transitions in
\cref{sec:estimating_phase_transitions_via_fits}, but found that this gave no
clear advantage.

Crucially, the VQE energy derivative and order parameters were both able to
correctly identify and locate phase 
transitions even when the overlap of the VQE state with the true ground state 
is very small. This hints that VQE may be able to give interesting 
information about a physical model, even when the possible circuit depths are 
not sufficient to prepare the ground state with high fidelity. But neither 
method works reliably for all models, making the situation analogous to that 
of trying to detect and locate phase transitions using existing techniques: 
It requires physical understanding of the model in question and experimentation
with different methods to get good results. Nevertheless, our work provides
evidence that near-term quantum computers and VQE could be used to understand
the phase diagrams of classically intractable models.


\section{Methods} 
\label{sec:methods}

In this section, we give a brief overview of the Variational Quantum Eigensolver 
and describe our ansatz circuits and the motivation behind them in more detail. 
We also found that reusing information obtained at different ansatz depths or 
different points in the phase diagram was crucial to keep the number of VQE 
runs tractable, and we will describe various methods to reuse this information to 
warm start optimisation.

\subsection{The variational method}
\label{sec:the-variational-method}

The Variational Quantum Eigensolver (VQE)~\cite{Peruzzo2014, McClean_2016} is a
prominent method for finding ground states $\ket{\psi_0}$ of quantum
Hamiltonians by classically optimising the parameters of a parametric unitary
$U(\vec\theta)$.
The overall goal of VQE is to minimize the objective function
\begin{equation}
  f(\vec\theta) = \braket{\psi(\vec\theta) | H | \psi(\vec\theta)}
       =\braket{\psi_i| U^\dagger(\vec\theta) H U(\vec\theta) |\psi_i},
\end{equation}
where $U(\vec\theta)$ is a family of unitaries parametrised by the classical
parameters $\vec\theta$, $\ket{\psi_i}$ an easily prepared initial state, and $H$
the Hamiltonian whose ground state $\ket{\psi_0}$ we wish to prepare.
By the variational principle $f(\vec\theta)$ is lower bounded by the ground 
state energy and can reach that bound only if $U(\vec \theta)$ is sufficiently
expressive and there exists some $\vec\theta^*$ such that 
$\ket{\psi(\vec\theta^*)} = \ket{\psi_0}$ for a ground state $\ket{\psi_0}$.
We will call $\ket{\psi(\vec\theta_{\textnormal{min}})}$ the \emph{VQE state}, 
where $\vec\theta_{\textnormal{min}}$ are the parameters found with minimal 
$f(\vec\theta)$. Due to the existence of many local minima, this may not be 
the global minimum but only the best parameters found with the methods described 
in \cref{sec:warm_starting_optimisation_with_previous_results}.

As parametric circuits $U(\vec\theta)$ we use the circuits generated by the 
Hamiltonian Variational ansatz (HV)~\cite{Wecker2015}. We split all 
Hamiltonians as 
\begin{equation}
  H = \sum_{i=1}^k H_i = \sum_{i=1}^k \sum_{j} H_{ij}
  \label{eq:hv_ansatz_hamiltonian_splitting}
\end{equation}
where each $H_i$ consists of mutually commuting, local terms $H_{ij}$. E.g. 
for the transverse field Ising model one could choose $H_1 = J \sum_{i=1}^{N-1} Z_i Z_{i+1}$
and $H_2 = h_x \sum_{i=1}^N X_i$ and the local terms would be $H_{1j} = J Z_j Z_{j+1}$ 
and $H_{2j} = h_x X_j$. The variational circuit of depth $p$ consists then
of $p$ identical layers with $k$ parameters each, takes the form
\begin{equation}
  U(\vec\theta) = \prod_{l=1}^p U_{\textnormal{layer}} (\vec\theta_l) 
  \quad \textnormal{where} \quad
  U_{\textnormal{layer}}(\vec\theta_l) = \prod_{i=1}^k e^{i \theta_{li} H_i}.
  \label{eq:hv_ansatz_circuit}
\end{equation}
and has $p k$ variational parameters.
If the initial state $\ket{\psi_0}$ is the ground state of one of the $H_i$ 
and $p$ is chosen sufficiently large, this variational circuit can represent
Trotterised, adiabatic annealing from this $H_i$ to the full Hamiltonian 
$H$. This means, that for sufficiently large $p$ (and some mild technical
constraints on the $H_i$) this circuit is guaranteed to be able to produce the
ground state of $H$ by virtue of the adiabatic theorem of quantum mechanics.
However, note that usually VQE is used with relatively small $p$, where the
adiabatic theorem does not hold.

\subsection{Warm starting optimisation with previous results} 
\label{sec:warm_starting_optimisation_with_previous_results}
The cost functions landscapes in VQE usually have many different local minima 
and finding the global one is hence hard. In all our experiments we 
want to run VQE for many different ansatz depths $p$ and at many different 
points in the phase diagram. This makes it possible to warm start the VQE 
optimisation process by reusing information obtained in previous runs at 
lower depths or other points in the phase diagram. 

Mele et al.~\cite{SmoothParameters_2022} note that the optimal VQE 
parameters $\theta_{li}^*$ at fixed $i$ often form a smooth function 
as a function of $l$. This means one can use good parameters found 
at low $p$ and warm start optimisation for higher $p$ by using a spline 
interpolation from the parameters found at low $p$. However, we found 
this to be true for some of the models we studied, but not all. 
For models where it does not hold, one may still warm start the optimisation
reusing the optimal parameters found at low $p$ and padding them with 
zeros to the required length at larger $p$.

Similarly, one may also want to reuse information obtained at other points 
in the phase diagram to warm start optimisation. 
When studying phase diagrams the family of Hamiltonians of interest is of the form 
\begin{equation}
  H(\vec g) = \sum_i c_i (\vec g) H_i
\end{equation}
with Hamiltonian parameters $\vec g$ and simple classical functions $c_i (\vec g)$.
This makes the VQE cost function
\begin{equation}
  f(\vec\theta; \vec g) = \braket{\psi(\vec\theta) |H(\vec g)| \psi(\vec\theta)} 
                        = \sum_i c_i(\vec g) \braket{\psi(\vec\theta) |H_i| \psi(\vec\theta)}.
\end{equation}
So if we evaluate each $\braket{\psi(\vec\theta) |H_i| \psi(\vec\theta)}$ 
in a separate run we can easily calculate $f(\vec\theta; \vec g)$ for all values
of $\vec g$ at that $\vec\theta$. We used this property to warm start VQE 
at a given $\vec g$ by using the optimal $\vec\theta$ observed in previous runs 
at different $\vec g$. Self et al.~\cite{Self_2021} go even further 
and propose running multiple instances of VQE at different $\vec g$ in parallel
and sharing information between them. But because we use LBFGS instead of Bayesian
optimisation as a classical optimiser this information sharing was not possible 
in our case.

The effect of these two methods to find better initial parameters and reuse 
already available information is demonstrated in
\cref{sec:the_effect_of_parameter_extrapolation}. While they worked 
well to make $\EVQE$ monotonously decay as a function of 
ansatz depth $p$ and smoother as a function of the Hamiltonian parameters, 
we strongly believe that---in particular for large $p$---they are still not
sufficient to find the global minimum. Especially, derived
quantities like $\frac{\Delta \EVQE}{\Delta p}$ where still noisy as 
the VQE energy improves only by very little for large $p$ and small deviations 
from the global minimum have a large impact here.

\section{Results} 
\label{sec:results}

For all models, we carried out extensive numerical simulations of VQE. 
For this work the focus was on investigating the optimal case performance of 
VQE to detect phase transition, so we leave the challenges associated with optimising VQE 
parameters on real hardware due to noise for future work. Hence we computed exact 
expectation values $f(\vec\theta) = \braket{\psi(\vec\theta)| H |\psi(\vec\theta) }$
from the full state vector and also exact, analytical gradients $\nabla f(\vec\theta)$
using \texttt{Yao.jl}'s~\cite{YaoFramework2019} automatic differentiation 
algorithms. Simulations of the 2D SSH model on up to 100 qubits were carried out using
\texttt{FLOYao.jl}~\cite{FLOYao_2022}, a fermionic linear optics backend for
\texttt{Yao.jl} based on the classical poly-time and space algorithm for 
simulating fermionic linear optics circuits~\cite{Terhal_2002,Bravyi_2012}.
For each model, we ran an initial $5-10$ VQE instances with random parameters 
initialized uniformly in $[0, 1/p]$. This data was then used to do two 
more runs with warm starting the optimisation with the previously found 
parameters, as described in \cref{sec:warm_starting_optimisation_with_previous_results}. 
The first run was warm started with initial parameters extrapolated from better 
lower depth parameters, if we happened to have found such. The second and final 
extra run was warm started with initial parameters from a better run at
different Hamiltonian parameters, if such a run happened. Unless otherwise
specified, all data in this section is from this final run, since it achieved the
lowest energy expectation value among all runs. For comparison, we also 
computed the exact ground state using the Lanczos algorithm implementation in 
\texttt{Arpack} for the TFIM and the bilinear-biquadratic chain and using full 
diagonalisation of the single particle Hamiltonian for the 2D SSH model. The 
overlaps between these exact ground states and the states found by VQE 
are presented in \cref{sec:more_data} for comparison. For the 2D TFIM and the 
bilinear-biquadratic chain these overlaps mark the phase transitions even more 
clearly than the order parameters or VQE energy derivatives. But since these 
overlaps would not be accessible in a real experiment, we did not include them 
in the main part.

\subsection{The transverse field Ising model}
\label{sec:the_tfim}
The Hamiltonian of the transverse field Ising model (TFIM) on a graph $G = (V, E)$
is  
\begin{equation}
  H_{\textnormal{TFIM}} = J \sum_{(i,j) \in E} Z_i Z_j + h_x \sum_{i \in V} X_i + h_z \sum_{i \in V} Z_i
  \label{eq:tfim_hamiltonian}
\end{equation}
where $Z_i$ and $X_i$ are the spin-1/2 operators in $z$- and $x$-direction
respectively. Throughout this paper, we set w.l.o.g. $J=-1$, $h_x \geq 0$ and
fix the bias field strength to be $h_z = \frac{1}{|V|^2}$. This means that the
only free parameter is the transverse field strength $h_x$ (or equivalently the
ratio $h_x/J$). The addition of $h_z$ breaks the $\mathbb{Z}_2$-symmetry and
ensures that even for finite chains the ground state in the ordered phase is a
simple product state---and not the unphysical superposition 
$\ket{0 \cdots 0} + \ket{1 \cdots 1}$---while simultaneously
only minimally altering the physics of the system because the gap to the higher 
energy states scales as $|V|^{-1}$.
As a simple product state, this initial state can be prepared easily in depth 1.

\begin{figure}
  \begin{center}
    \begin{tikzpicture}[very thick]


  \begin{scope}
    \foreach \xx/\x/\phi in {0/0/80,1/1.5/120,2/3/70,3/5/125,4/6.5/100}{
        \node[inner sep=1ex] (\xx) at (\x,0) {};
        \shade[ball color=gray] (\xx) circle (1ex);
    }

    \draw[color=c0] (0) -- (1);

    \foreach \a/\b in {1/2,3/4}{
      \draw[color=gray] (\a) -- (\b);
    }

    \draw[dotted,color=c0] ($(2) + (0.3,0)$) -- ($(3) - (0.3,0)$);

    \node[color=c0] at (0.7,0.4) {$J Z_1 Z_2$};
    \node[color=c0] at (4.,0.4) {$J Z_i Z_{i+1}$};

    \node (slabel) at (2.2, 1) {$S=\frac{1}{2}$};
    \draw[-stealth,ultra thin] (slabel) to[out=220,in=70] (1);
    \draw[-stealth,ultra thin] (slabel) to[out=330,in=110] (2);
  \end{scope}

  \begin{scope}[shift={(0, -1.5)}]
    \foreach \xx/\x/\phi in {0/0/80,1/1.5/120,2/3/70,3/5/125,4/6.5/100}{
        \node[inner sep=1ex] (\xx) at (\x,0) {};
        \shade[ball color=gray] (\xx) circle (1ex);
    }

    \draw[color=gray] (0) -- (1);

    \foreach \a/\b in {1/2,3/4}{
      \draw[color=c1] (\a) -- (\b);
    }

    \draw[dotted,color=gray] ($(2) + (0.3,0)$) -- ($(3) - (0.3,0)$);

    \node[color=c1] at (2.2,0.4) {$J Z_{2} Z_3$};
    \node[color=c1] at (5.7,0.4) {$J Z_{L-1} Z_L$};
  \end{scope}

  \begin{scope}[shift={(0, -3.0)}]
    \foreach \xx/\x/\phi in {0/0/80,1/1.5/120,2/3/70,3/5/125,4/6.5/100}{
        \draw[-latex,color=c2] ($(\x, 0) - (-135:0.3cm)$) -- ($(\x, 0) + (-135:0.5cm)$) ;
        \node[inner sep=1ex] (\xx) at (\x,0) {};
        \shade[ball color=gray] (\xx) circle (1ex);
    }

    \draw[color=gray] (0) -- (1);

    \foreach \a/\b in {1/2,3/4}{
      \draw[color=gray] (\a) -- (\b);
    }

    \draw[dotted,color=gray] ($(2) + (0.3,0)$) -- ($(3) - (0.3,0)$);

    \foreach \x/\l in {0.6/$h_x X_1$,2.1/$h_x X_2$,3.7/$h_x X_3$,6.0/$h_x X_{L}$}{
      \node[color=c2] at (\x,0.3) {\l};
    }
  \end{scope}

  \begin{scope}[shift={(0, -4.5)}]
    \foreach \xx/\x/\phi in {0/0/80,1/1.5/120,2/3/70,3/5/125,4/6.5/100}{
        \draw[-latex,color=c3] ($(\x, 0) - (90:0.4cm)$) -- ($(\x, 0) + (90:0.6cm)$) ;
        \node[inner sep=1ex] (\xx) at (\x,0) {};
        \shade[ball color=gray] (\xx) circle (1ex);
    }

    \draw[color=gray] (0) -- (1);

    \foreach \a/\b in {1/2,3/4}{
      \draw[color=gray] (\a) -- (\b);
    }

    \draw[dotted,color=gray] ($(2) + (0.3,0)$) -- ($(3) - (0.3,0)$);

    \foreach \x/\l in {0.6/$h_z Z_1$,2.1/$h_z Z_2$,3.7/$h_z Z_3$,6.0/$h_z Z_{L}$}{
      \node[color=c3] at (\x,0.3) {\l};
    }
  \end{scope}
\end{tikzpicture}
  \end{center}
  \caption{Sketch of the 1D TFIM Hamiltonian and its splitting into sub-Hamiltonians.
    For the ansatz circuits in 
    \cref{eq:hv_ansatz_hamiltonian_splitting} we decompose it into the $ZZ$ 
    interaction on all {\color{c0} odd edges}, the $ZZ$ interaction on all 
    {\color{c1} even edges}, the {\color{c2} transverse field} and the 
    {\color{c3} bias field}.
    For the 2D TFIM the same decomposition applies, but instead of splitting 
    the interaction part into 2 it is split into 4 parts according to the same 
    edge colouring also used in \cref{fig:ssh2d-sketch}. 
  }
  \label{fig:tfim-sketch}
\end{figure}
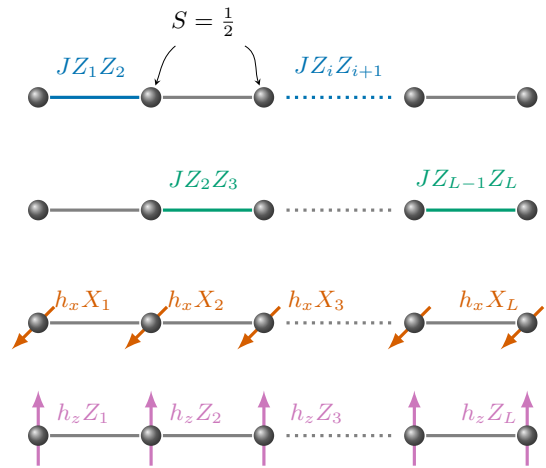

For $|h_x| \ll |J|$ the ground state of the TFIM is the simple product state 
$\ket{\psi_{\textnormal{ordered}}} = \ket{1 \cdots 1}$ and 
the TFIM is said to be in the ordered phase. This is also the state we 
use as an initial state for VQE with the HV ansatz.
For $|h_x| \gg |J|$ the ground state is similarly simple:
$\ket{\psi_{\textnormal{disordered}}} = \ket{+ \cdots +}$ and the system is said
to be in the disordered phase.
In 1D, the Kramers-Wannier Duality~\cite{Fradkin_1978} can be used to show that 
the phase transition is exactly at $\left|\frac{h_x}{J}\right| = 1$.

As an order parameter to witness the phase transition in the TFIM we use 
the $z$-magnetisation 
\begin{equation}
  m_z = \frac{1}{N} \sum_{\textnormal{sites } i} Z_i
  \label{eq:z_magnetization}
\end{equation}
which vanishes in the disordered phase and is $\pm 1$ in the ordered limit,
$-1$ with our choice of a positive bias field strength $h_z$. In the
thermodynamic limit, its derivative with respect to the field strength $h_x$ is
known to diverge at the phase transition.

\begin{figure}
  \begin{center}
    \input{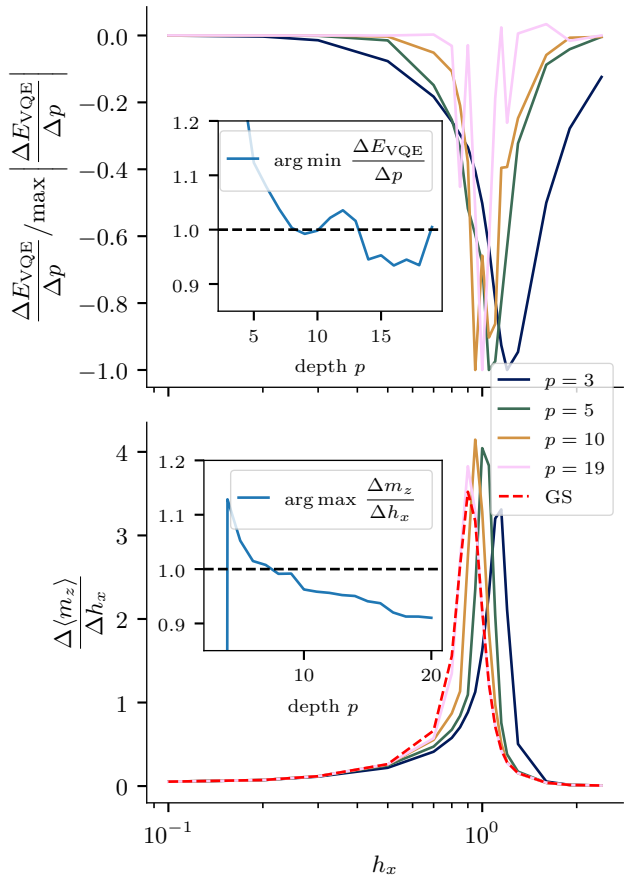}
  \end{center}
  \caption{Predicting the phase transition of the 1D TFIM on 16 sites via the
    measure of the VQE energy derivative (top) and the derivative of the
    $z$-magnetization (bottom). The insets show the argmin (argmax) of the
    respective quantities with respect to the transverse field strength $h_x$
    as a function of ansatz depth $p$. In the lower plot the results for the
    exact ground state found are shown with red dashed lines.
  }
  \label{fig:tfim_phase_transition_prediction}
\end{figure}

\begin{figure}
  \begin{center}
    \input{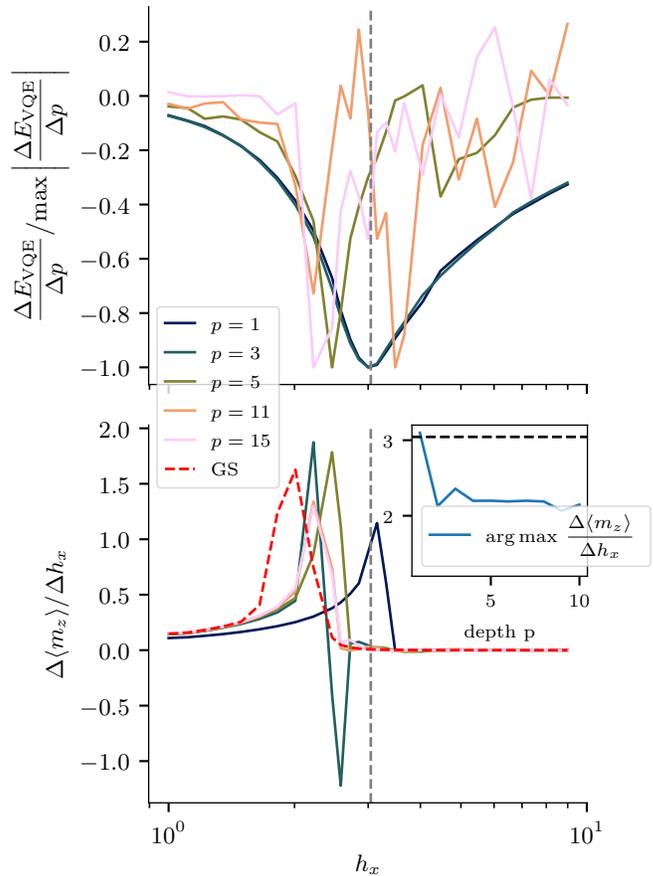}
  \end{center}
  \caption{Predicting the phase transition of the $5 \times 5$ TFIM via the measure of
    the VQE energy derivative (top) and the derivative of the $z$-magnetization
    (bottom). The grey vertical lines are at $h_x = 3.044$, the phase transition
    point in the thermodynamic limit \cite{deJongh_1998,Bloete_2002}. In the 
    lower plot the results for the exact ground state found are shown with red
    dashed lines.
  }
  \label{fig:2dtfim_phase_transition_prediction}
\end{figure}

In~\cref{fig:tfim_phase_transition_prediction,fig:2dtfim_phase_transition_prediction} 
we compare our two methods of predicting phase transitions from VQE results for
the TFIM. In the 1D case, we see that both the derivative of the
$z$-magnetisation and the VQE energy derivative correctly identify the phase
transition and locate it near $h_x = 1$ by having a pronounced peak there. Both
methods overestimate the location of the phase transition for small $p$ and
converge to lower values with increasing $p$.
However, the argmax of the derivative of
the magnetization converges to a value below $h_x = 1$---most likely due
to finite size effects---while the argmin of the VQE energy derivative does not
converge to a clear value, most likely due to the derivatives being
vanishingly small for large $p$ and hence hard to compute using finite
differences from noisy values. The multiple minima for $p=19$ in the VQE energy
derivatives are most likely spurious and due to the fact that we used
finite differences on noisy data to compute the derivatives.

In the 2D case, the
VQE energy derivative works well for very low $p$ but is inconclusive for
larger $p$ due to the challenges of finding the global minimum of the cost
function and the VQE energy derivative being sensitive to this. Unsurprisingly,
the derivative of the $z$-magnetisation works better for estimating the phase
transition at higher $p$. However, due to finite size effects its maximum is
not at the point it would be in the thermodynamic
limit~\cite{deJongh_1998,Bloete_2002}.
At first, it might appear surprising that the VQE energy 
derivative behaves so similarly before and after the phase transition (remember 
that the initial state is the $h_x = 0$ ground state). But it turns out 
that evolution with $\sum_i X_i$ and $\sum_i Z_i$ 
can be combined to apply $R_y(\pi / 2)$ in the first VQE layer, taking 
the ground state at $h_x = 0$ to the ground state at $J = 0 = h_z$.
In \cref{sec:more_tfim_data} we also present the fidelity of the 
VQE state with the exact ground state for all available data points as well 
as all data going into the top subplot of
\cref{fig:tfim_phase_transition_prediction,fig:2dtfim_phase_transition_prediction} 
to show the vanishingly small energy derivatives for larger $p$ more clearly.

\subsection{The bilinear-biquadratic chain}
\label{sec:the_bb_chain}

The Hamiltonian of the bilinear-biquadratic chain on $L$ sites is 
\begin{equation}
  H_{\textnormal{BBC}} = \cos(\phi) \sum_{i}^{L-1} \vec S_i \cdot \vec S_{i+1} 
                       + \sin(\phi) \sum_{i}^{L-1} (\vec S_i \cdot \vec S_{i+1})^2
  \label{eq:bb_hamiltonian}
\end{equation}
where $S_i^{(\alpha)}$ is the spin-1 operator in $\alpha$-direction on site $i$.
The only free parameter in this model is the angle $\phi$. 

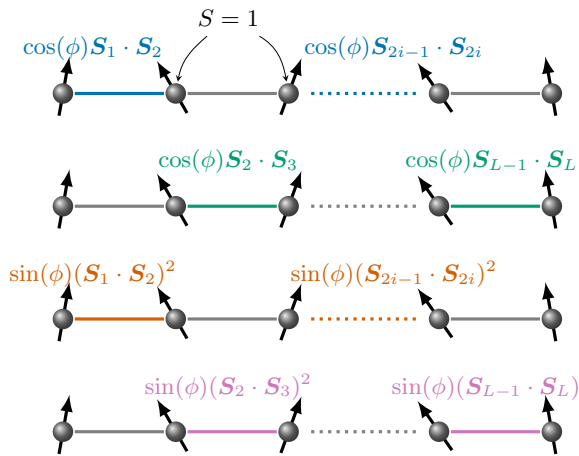
\begin{figure}
  \begin{center}
    \begin{tikzpicture}[very thick]
  \begin{scope}
    \foreach \xx/\x/\phi in {0/0/80,1/1.5/120,2/3/70,3/5/125,4/6.5/100}{
        \draw[-latex] ($(\x, 0) - (\phi:0.3cm)$) -- ($(\x, 0) + (\phi:0.5cm)$) ;
        \node[inner sep=1ex] (\xx) at (\x,0) {};
        \shade[ball color=gray] (\xx) circle (1ex);
    }

    \draw[color=c0] (0) -- (1);

    \foreach \a/\b in {1/2,3/4}{
      \draw[color=gray] (\a) -- (\b);
    }

    \draw[dotted,color=c0] ($(2) + (0.3,0)$) -- ($(3) - (0.3,0)$);

    \node[color=c0] at (0.4,0.6) {$\cos(\phi) \vec{S}_1 \cdot \vec{S}_2$};
    \node[color=c0] at (4.4,0.6) {$\cos(\phi) \vec{S}_{2i-1} \cdot \vec{S}_{2i}$};

    \node (slabel) at (2.2, 1) {$S=1$};
    \draw[-stealth,ultra thin] (slabel) to[out=220,in=80] (1);
    \draw[-stealth,ultra thin] (slabel) to[out=330,in=100] (2);
  \end{scope}

  \begin{scope}[shift={(0,-1.5)}]
    \foreach \xx/\x/\phi in {0/0/80,1/1.5/120,2/3/70,3/5/125,4/6.5/100}{
        \draw[-latex] ($(\x, 0) - (\phi:0.3cm)$) -- ($(\x, 0) + (\phi:0.5cm)$) ;
        \node[inner sep=1ex] (\xx) at (\x,0) {};
        \shade[ball color=gray] (\xx) circle (1ex);
    }

    \draw[color=gray] (0) -- (1);

    \foreach \a/\b in {1/2,3/4}{
      \draw[color=c1] (\a) -- (\b);
    }

    \draw[dotted,color=gray] ($(2) + (0.3,0)$) -- ($(3) - (0.3,0)$);

    \node[color=c1] at (2.2,0.6) {$\cos(\phi) \vec{S}_{2} \cdot \vec{S}_3$};
    \node[color=c1] at (5.7,0.6) {$\cos(\phi) \vec{S}_{L-1} \cdot \vec{S}_L$};
  \end{scope}

  \begin{scope}[shift={(0,-3)}]
    \foreach \xx/\x/\phi in {0/0/80,1/1.5/120,2/3/70,3/5/125,4/6.5/100}{
        \draw[-latex] ($(\x, 0) - (\phi:0.3cm)$) -- ($(\x, 0) + (\phi:0.5cm)$) ;
        \node[inner sep=1ex] (\xx) at (\x,0) {};
        \shade[ball color=gray] (\xx) circle (1ex);
    }

    \draw[color=c2] (0) -- (1);

    \foreach \a/\b in {1/2,3/4}{
      \draw[color=gray] (\a) -- (\b);
    }

    \draw[dotted,color=c2] ($(2) + (0.3,0)$) -- ($(3) - (0.3,0)$);

    \node[color=c2] at (0.4,0.6) {$\sin(\phi) (\vec{S}_1 \cdot \vec{S}_2)^2$};
    \node[color=c2] at (4.4,0.6) {$\sin(\phi) (\vec{S}_{2i-1} \cdot \vec{S}_{2i})^2$};
  \end{scope}

  \begin{scope}[shift={(0,-4.5)}]
    \foreach \xx/\x/\phi in {0/0/80,1/1.5/120,2/3/70,3/5/125,4/6.5/100}{
        \draw[-latex] ($(\x, 0) - (\phi:0.3cm)$) -- ($(\x, 0) + (\phi:0.5cm)$) ;
        \node[inner sep=1ex] (\xx) at (\x,0) {};
        \shade[ball color=gray] (\xx) circle (1ex);
    }

    \draw[color=gray] (0) -- (1);

    \foreach \a/\b in {1/2,3/4}{
      \draw[color=c3] (\a) -- (\b);
    }

    \draw[dotted,color=gray] ($(2) + (0.3,0)$) -- ($(3) - (0.3,0)$);

    \node[color=c3] at (2.2,0.6) {$\sin(\phi) (\vec{S}_{2} \cdot \vec{S}_3)^2$};
    \node[color=c3] at (5.7,0.6) {$\sin(\phi) (\vec{S}_{L-1} \cdot \vec{S}_L)^2$};
  \end{scope}

\end{tikzpicture}
  \end{center}
  \caption{Sketch of the bilinear-biquadratic chain Hamiltonian. For the ansatz 
    circuits we decompose \cref{eq:bb_hamiltonian} into the 
    {\color{c0} linear interaction on odd edges},
    the {\color{c1} linear interaction on even edges}, 
    the {\color{c2} quadratic interaction on odd edges}
    and the {\color{c3} quadratic interaction on even edges}.
  }
  \label{fig:bbc-sketch}
\end{figure}

One checks readily that the terms $\vec S_i \cdot \vec S_{i+1}$ (and hence 
also the terms $(\vec S_i \cdot \vec S_{i+1})^2$) commute with the total 
spin in $z$ direction 
\begin{equation}
  S_{\mathrm{total}}^z = \sum_i S_i^z
\end{equation}
and by $SO(3)$ symmetry the same is true for the analogously defined
$S_{\mathrm{total}}^x$ and $S_{\mathrm{total}}^y$. This makes
$S_{\mathrm{total}}^\alpha$ ($\alpha = x, y, z$) conserved quantities of our
ansatz circuits.

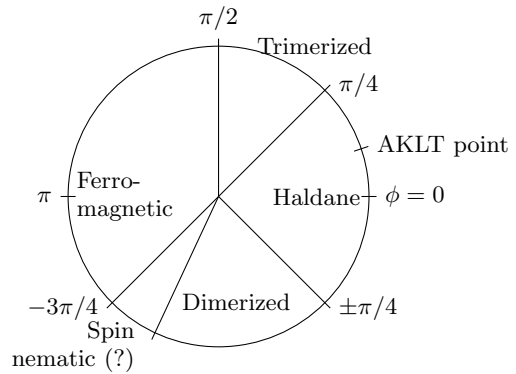
\begin{figure}
  \begin{center}
    \begin{tikzpicture}[scale=2]
  \draw (0,0) circle (1);
  \draw (18.5:0.95) -- (18.5:1.05) node[pos=1,right] {AKLT point};
  \draw (0,0) -- (45:1.05) node[pos=1,right] {$\pi/4$};
  \draw (0,0) -- (90:1.05) node[pos=1,above] {$\pi/2$};
  \draw (0,0) -- (-135:1.05) node[pos=1,left] {$- 3\pi/4$};
  \draw (0,0) -- (-115:1.05);
  \draw (0,0) -- (-45:1.05) node[pos=1,right] {$\pm\pi/4$};

  \draw (0.95,0) -- (1.05,0) node[pos=1,right] {$\phi=0$};
  \draw (-0.95,0) -- (-1.05,0) node[pos=1,left] {$\pi$};

  \node[left] at (1,0) {Haldane};
  \node[right] at (0.2,1.0) {Trimerized};
  \node[right,align=left] at (-1,0) {Ferro-\\magnetic};
  \node[left,align=right] at (-0.5,-1) {Spin \\ nematic (?)};
  \node[right] at (-0.3,-0.7) {Dimerized};

\end{tikzpicture}
  \end{center}
  \caption{The phase diagram of the bilinear-biquadratic chain. The boundaries
    of the Haldane, trimerised, ferromagnetic can be exactly calculated, but the 
    existence of a spin nematic phase in $-3 \pi / 4 < \phi \lesssim -0.67 \pi$ is still
    disputed~\cite{Laeuchli_2006,Buchta_2005}.
  }
  \label{fig:bbc_phase_diagram}
\end{figure}

In \cref{fig:bbc_phase_diagram} we show the phase diagram of the
bilinear-biquadratic chain as given in~\cite{Laeuchli_2006,Buchta_2005}. At the
AKLT point $\phi = \arctan (1/3)$ the ground state is exactly expressible as a
matrix product state with bond dimension two, the AKLT state~\cite{AKLT_1987},
and is four-fold degenerate due to the two spin-1/2 degrees of freedom at each
end of the open boundary AKLT state. We use the AKLT state as the initial state for VQE 
and to enable our ansatz to produce states with different
$S_{\mathrm{total}}^\alpha$ ($\alpha = x, y, z$) we include the two spin-1/2
degrees of freedom at either end of the chain in the variational parameters.
As an MPS with bond dimension 2 the AKLT state can be prepared in linear 
depth with the use of one ancilla qubit~\cite{Schoen_2005} or using its 
symmetries and fusion measurements even in constant depth~\cite{Smith_2022}.

\begin{figure}
  \begin{center}
    \input{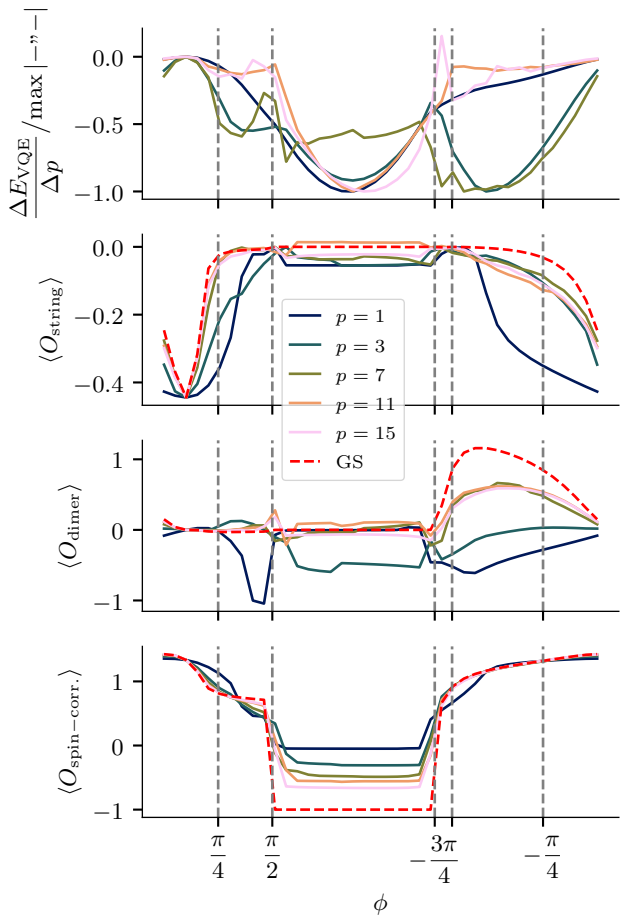}
  \end{center}
  \caption{
    Comparison of the VQE energy derivative (top), string order (second subplot),
    dimerisation (third subplot) and spin correlation (bottom)  as a function of
    $\phi$ for different VQE depths $p$ in the 12 qutrit bilinear-biquadratic
    chain. The grey vertical lines denote the exactly known phase transitions in
    the thermodynamic limit. For comparison, we also show the results 
    for the exact ground state with red dashed lines.
  }
  \label{fig:bilbiq_phase_transition_prediction}
\end{figure}

There are different order parameters to detect the different phases of the 
bilinear-biquadratic chains. The Haldane phase is heralded by the non-vanishing 
of the \emph{string order} parameter 
\begin{equation}
  O_{\textnormal{string}}^\alpha 
  = \lim_{|j-i| \to \infty} S_i^\alpha e^{i\pi (S_{i+1}^\alpha + \cdots + S_{j-1})}   S_j^\alpha
  \label{eq:string_order}
\end{equation}
with $\alpha = x, y, z$. Due to the symmetry of the Hamiltonian, it does not 
matter which $\alpha$ one chooses and we used $\alpha = z$ throughout. 
For the data shown in \cref{fig:bilbiq_phase_transition_prediction}  
we used $i=2$ and $j=L-1$ as the best approximation of ``in the bulk'' we can do 
with the small systems we are able to simulate.
The dimerisation of the chain is measured by the \emph{staggered dimerisation}
\begin{equation}
  O_{\textnormal{dimer}} = \frac{1}{L-2} \sum_{i=2}^{L-1} (-1)^i (\vec S_{i-1} \cdot \vec S_i - \vec S_{i} \cdot \vec S_{i+1})
  \label{eq:dimerisation}
\end{equation}
which vanishes outside the dimerised phase and is $2$ for a perfectly dimerised
state. 
The ferromagnetic phase is indicated simply by the \emph{spin correlation}
\begin{equation}
  O_{\textnormal{spin-corr}} = \frac{1}{L-1} \sum_{i=2}^L  \vec S_{i-1} \cdot \vec S_i
  \label{eq:spin_correlation}
\end{equation}
which again takes the value 2 for a perfectly ferromagnetic state. 

Our numerical results for the different techniques of detecting phase transitions 
for the 12-qutrit BBC are compared in \cref{fig:bilbiq_phase_transition_prediction}. 
In the top subplot, we show the VQE energy derivative, again normalized to have
a maximum absolute value of 1 for all different $p$. In the three lower subplots 
we show the string order \cref{eq:string_order}, the staggered dimerisation 
\cref{eq:dimerisation} and the spin correlation \cref{eq:spin_correlation}
for the same different ansatz depths $p$ and the exact ground state.
Compared to the TFIM and \cref{fig:tfim_phase_transition_prediction} the 
situation is less clear for the bilinear-biquadratic chain. The spin-correlation 
signals the ferromagnetic phase well, even for low $p$ and even 
though the true ground state can never be reached using our VQE ansatz (cf. 
\cref{fig:bilbiq_vqe_fidelities}) due to the conserved quantities of the ansatz.
The existence of a Haldane phase is correctly detected by the string order 
for $p \geq 7$ here, similar to the dimerised phase which is detected 
by the staggered dimerisation. But in both cases, the exact point of the phase 
transition cannot be detected, partially due to finite size effects as shown 
by the data for the exact ground state. The VQE energy derivative marks some 
of the phase transitions with discontinuities but behaves differently depending 
on the VQE depth. For some $p$ the VQE energy derivative correctly
signals the ferromagnetic phase, for others the dimerised or trimerised phase
and the AKLT point is clearly marked as the point where the VQE energy stays
constant, simply because here the initial state is already the ground state.
However, the phase boundaries of the AKLT phase are not clearly marked by the 
VQE energy derivative. This makes sense because the AKLT phase is a topological 
phase and as such, it cannot be detected by local observables. The situation 
around $\phi = - \frac{3 \pi}{4}$ and whether a phase between the 
ferromagnetic and the dimerised phase exists is unclear from all our methods.
For comparison, we also show the VQE energy derivative for all depths $p$,
parameters $\phi$ and simulated system sizes $L$ in \cref{sec:more_bilbiq_data}.

\subsection{The 2D SSH model}
\label{sec:the_2d_ssh_model}

\begin{figure}
  \begin{center}
    \begin{tikzpicture}

  \draw[fill,color=gray!50!white,rounded corners=2pt,text=black] (4.2,4.2) rectangle (5.8, 5.8)
    node[pos=0.5,align=center] {unit\\cell};

  \foreach \x in {0,1,2,3.5,4.5,5.5}
    \foreach \y in {0,4.5}
      {
        \draw[very thick,color=c0] (\y,\x) -- (\y+1, \x);
        \draw[very thick,color=c2] (\x,\y) -- (\x, \y+1);
      }

  \foreach \x in {1,3.5}
    \foreach \y in {0,1,2,3.5,4.5,5.5}
      {
        \draw[thin,color=c1] (\x,\y) -- (\x+1, \y);
        \draw[thin,color=c3] (\y,\x) -- (\y, \x+1);
      }

  \foreach \i in {0,1,2,3.5,4.5,5.5}
  {
    \draw[very thick,dotted,color=c0] (2.5,\i) -- (3,\i);
    \draw[very thick,dotted,color=c2] (\i,2.5) -- (\i,3);
  }

  \foreach \i/\x in {0/0,1/1,2/2,3/3.5,4/4.5,5/5.5}
    \foreach \j/\y in {0/0,1/1,2/2,3/3.5,4/4.5,5/5.5}
      {
        \node[dot,gray] (\i,\j) at (\x,\y) {};
      }

  \foreach \x/\y/\c/\l in {0.2/5./c2/$v$,
                           1.2/4./c3/$w$,
                           0.5/5.3/c0/$v$,
                           1.5/4.3/c1/$w$,
                           0.5/4.7/c0/$v$,
                           1.5/3.7/c1/$w$,
                           0.8/5.0/c2/$v$,
                           1.8/4.0/c3/$w$}
      {
        \node[color=\c] at (\x, \y){\l};
      }

  \foreach \x/\y in {0/0,  5.5/5.5}
      {
        \node[fill,circle,c4,inner sep=1pt,text=black] at (\x, \y){$\mu$};
 
  \foreach \x/\y in {5.5/0, 0/5.5}
      {
        \node[fill,circle,c4,inner sep=-2pt,text=black] at (\x, \y){$-\mu$};
      }     }

\end{tikzpicture}
  \end{center}
  \caption{Sketch of the 2D SSH model. The intra cell (thick lines)
    hopping strength is $v$ while the inter cell (thin lines) hopping
    strength is $w$. On the top left and bottom right corner we added a
    negative potential and with strength $-\mu$ and on the top right and bottom
    left corner a positive potential with strength $\mu$ to make sure the ground
    state in the topological limit has occupation number 0 or 1 on the corners.
    Again, we decompose the Hamiltonian into five sub-Hamiltonians according 
    to the five colours we use for the different terms in this sketch to create 
    the ansatz circuits as in \cref{eq:hv_ansatz_hamiltonian_splitting}.
  }
  \label{fig:ssh2d-sketch}
\end{figure}
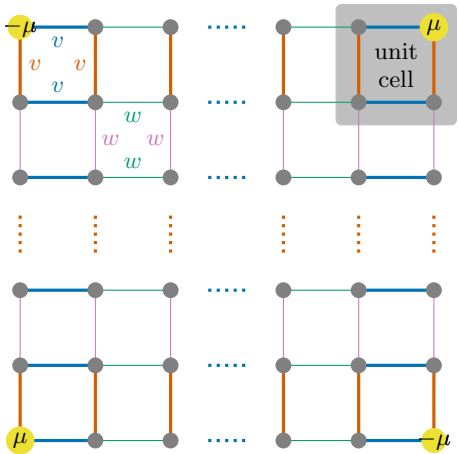

We use the following Hamiltonian for the 2D SSH model 
\begin{equation}
  \begin{aligned}
    H_{\textnormal{SSH}} &= -v \kern-1em \sum_{\substack{\textnormal{intercell} \\
                                          \textnormal{edges } \braket{i,j}}} \kern-1em 
                     c_i^\dagger c_j + c_j^\dagger c_i 
              -w \kern-1em \sum_{\substack{\textnormal{intracell} \\
                                 \textnormal{edges } \braket{i,j}}} \kern-1em 
                     c_i^\dagger c_j + c_j^\dagger c_i  \\
          &+ \mu (n_{L} + n_{L^2-L+1} - n_{1} - n_{L^2} )
  \end{aligned}
  \label{eq:2d_ssh_hamiltonian}
\end{equation}
where the $c_i$ are fermionic annihilation operators located on the vertices of 
an $L \times L$ square lattice that is subdivided into $2 \times 2$ unit cells 
as shown in \cref{fig:ssh2d-sketch}. The total number of electrons is $L^2/2$,
so the system is at half-filling. The second line are positive onsite 
potentials on the upper right and lower left sites and negative onsite 
potentials on the upper left and lower right site. Up to the onsite potentials
in the second line, this is the same model studied in~\cite{Obana_2019}.
Throughout the paper, we fixed the energy scale $v + w = 2$ and the onsite
potential strength to $\mu = \frac{1}{L^2}$. This means the only free parameter
is the hopping strength ratio $\frac{v}{w}$. The addition of $\mu$ ensures that
the ground state in the limit $\frac{v}{w} \to 0$ has definite occupation
numbers on all four corners while only minimally altering the physics because
the excited modes have a minimal energy scaling with $L^{-1}$.

In the limit $v \gg w$ the system is in the trivial phase and at half-filling 
the ground state is the product of putting two particles into the two 
lowest energy modes in each unit cell. The circuit depth to prepare this 
initial state depends on the encoding used to map fermions to qubits. Using 
the efficient Hamiltonian Variational ansatz from~\cite{Cade_2019} it 
can be prepared in depth $L$ using Givens rotations~\cite{Jiang_2018}. Using 
a local encoding~\cite{Klaassen_2021} the same scheme based on Givens rotations 
can be used to prepare the trivial ground state in constant depth.
In the limit $v \ll w$ the system is in the 
topological phase and the ground state is the product of putting two particles 
into the two lowest energy modes in each orange square in \cref{fig:ssh2d-sketch}. 
In absence of the onsite potential the four corners now host topologically 
protected zero modes, two of which will be filled. Because the salient
feature of the 2D SSH model---the
absence or presence of these zero modes---only plays a role at half filling we
only consider that case from now on.
Because we use the Jordan-Wigner transformation to map the fermionic system to 
qubits, we cannot directly use the ordering of \cref{eq:hv_ansatz_circuit} 
for the ansatz circuits. Instead, we use the ansatz circuits already used 
in~\cite{Cade_2019} to map the fermionic sites on a square
lattice to qubits laid out in a square lattice.

As an order parameter to detect the phase transition in the 2D SSH model, 
we use the \emph{corner occupation order parameter (COOP)}
\begin{equation}
  O_{\textnormal{COOP}} = e^{i \pi (n_{1} + n_{L} + n_{L^2-L+1} + n_{L^2})}.
  \label{eq:coop}
\end{equation}
In the trivial limit $v \gg w$ the occupation per site in the ground state at half filling 
is $\frac{1}{2}$ on all sites, whereas in the topological limit $v \ll w$ the four 
corners have occupation $0$ or $1$. This implies that $O_{\textnormal{COOP}} = 0$
in the trivial limit and $O_{\textnormal{COOP}} = 1$ in the topological limit.

\begin{figure}
  \begin{center}
    \input{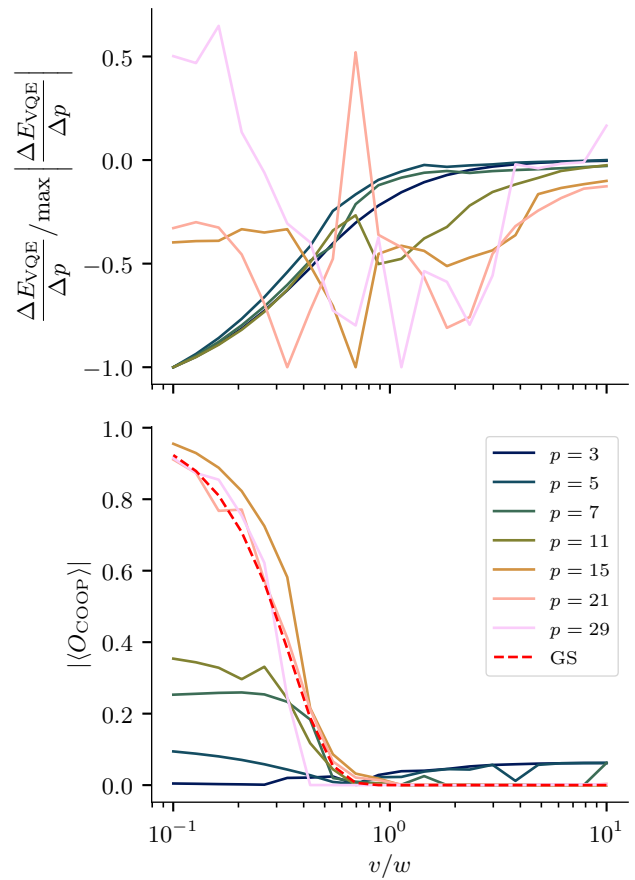}
  \end{center}
  \caption{
    Comparison of VQE energy derivative (top) and corner occupation order parameter
    (bottom) as a function of $v/w$ for different VQE depths $p$ in the $10
    \times 10$ SSH model. The complete, non-normalized data for the 
    VQE energy derivatives is also shown in \cref{fig:ssh_energy_derivatives}.
  }
  \label{fig:ssh_phase_transition_prediction}
\end{figure}

Our numerical findings for the 2D SSH model are summarized in
\cref{fig:ssh_phase_transition_prediction}. Again, we show the VQE energy
derivative in the top subplot and the order parameter---in this case, the corner
occupation order parameter \cref{eq:coop}---in  the bottom. In comparison to
the same data for the TFIM shown in \cref{fig:tfim_phase_transition_prediction}
the situation is quite different here: While the corner occupation order
parameter correctly predicts and locates the phase transition, albeit only for
larger ansatz depths $p$, the sympathetic eye may see a change in behaviour of
the VQE energy derivative for lower $p$ around the phase transition. For
large $p$ we mostly see numerical noise and the effects of not being able to
find the global minimum in the VQE energy derivative. The good agreement between
the corner occupation order
parameter in the exact ground state and the VQE states for $p \geq 15$ is
surprising---but encouraging in the context of this paper---given the very bad
fidelities shown in \cref{sec:more_ssh_data} between those two states.
The marked difference between the VQE energy derivative in the TFIM (shown 
at the top of \cref{fig:tfim_phase_transition_prediction}) and for the 
2D SSH is in part explained by the fact that for the TFIM we know how to 
take the $h_x = 0$ ground state (i.e. the initial state for VQE) 
to the $J = 0 = h_z$ ground state in one VQE layer (see last paragraph in 
\cref{sec:the_tfim}) whereas for the 2D SSH model, the different Chern 
numbers of the trivial and the topological phase mean that no translation 
invariant, low depth, non-interacting circuit exists that can take the ground state 
of one phase to the ground state of the other phase. See 
\cref{sec:vqe_hardness_in_noninteracting_free_fermion_systems} for a proof of 
this. This is also consistent with the fact that the VQE energy derivative is a 
local observable and as such not expected to be able to differentiate between 
the topological and trivial phase, similar to the way the VQE energy derivative 
didn't mark the boundaries of the AKLT phase in the bilinear-biquadratic chain 
clearly.

\section{Discussion and Outlook} 
\label{sec:disussion_and_outlook}

We have demonstrated the usefulness of low-depth VQE to map out phase diagrams.
Even when the circuit depth is not sufficient to prepare the true ground state 
with high fidelity, measuring order parameters or the VQE energy derivative 
is able to detect phase transitions in the systems simulated by us. Overall, 
measuring order parameters will require less precision in the estimation of 
observables than measuring the VQE energy derivative, since one needs the raw
expectation value of the order parameters and not the difference between two
(possibly very close) expectation values. 
However, considering the VQE energy has the marked advantage of being model 
agnostic and may prove useful to gain a rough overview of the phase diagram 
in question. In contrast to measuring order parameters, the VQE energy derivative 
also has the advantage that it gives more precise answers for lower depths where 
the energy decreases faster as a function of $p$ whereas for large $p$ noise 
makes it harder to accurately measure the VQE energy derivative. For order 
parameters, on the other hand, it is unclear if lower $p$ is advantageous 
because the light cones of local observables don't span the whole system 
and finite size effects are hence suppressed, or if larger $p$  give more 
accurate results because the VQE state has a larger overlap with the 
(finite size) ground state.
 
What remains open are questions of experimental feasibility. A quick 
back-of-the-envelope calculation shows that the relative accuracy 
$\left| \frac{\EVQE(p+1) - \EVQE(p)}{\EVQE(p)} \right|$ needed to estimate 
the VQE energy derivative accurately enough to accurately locate the phase
transition ranges from $10^{-4}$ (setting $p=10$ in the 16-site TFIM where
$|E_0| \sim 20$) to $10^{-3}$ (setting $p=10$ in the 12-qutrit
bilinear-biquadratic chain where $|E_0| \sim 30$). These required relative
accuracies were the same for an 8-site or 12-site TFIM and 8-qutrit
bilinear-biquadratic chain, hinting that they also remain the same for larger
system sizes. However, this is still one order of magnitude away from the
relative accuracies of $10^{-2}$ achieved with the help of sophisticated error
mitigation schemes in~\cite{Stanisic_2022,Yu_2022}.

This is in stark contrast 
to the situation where one already knows what phases are expected and is able 
to measure the corresponding order parameter. Now it is often, but not always,
sufficient to distinguish between the order parameter being zero outside of the 
phase of interest, and having a non-zero value inside the phase of interest.
Depending on the critical exponent, the order parameter will also very 
quickly increase close to the phase transition. E.g.\ in the case of 1D TFIM 
we have $\braket{m_z} \sim \left(\frac{h_x}{J} - 1\right)^{\frac{1}{8}}$ meaning that to 
locate the phase transition up to precision $\epsilon$ we need to learn 
$\braket{m_z}$ only up to precision $\epsilon^{\frac{1}{8}}$. And the phase 
transitions to the ferromagnetic phase (certainly from the trimerised phase) in
the bilinear-biquadratic chain are of first order, meaning there is a finite
discontinuity in $\braket{O_{\textnormal{spin-corr.}}}$ as a function of $\phi$
that is easy to detect, even from noisy data.

It remains unclear, how the depth needed to accurately locate the phase 
transitions will scale with the system size. The data shown in~\cref{fig:bilbiq_vqe_hardness}
and similar analysis for the 1D TFIM and 2D SSH model indicate that to see 
similar behaviour in the VQE energy derivative the needed circuit depth scales 
with the system size. On the other hand the analysis done in~\cite{Dreyer_2021} 
and~\cref{sec:vqe_hardness_in_noninteracting_free_fermion_systems} show that 
in the thermodynamic limit for non-interacting fermionic systems only constant 
circuit depth is required to prepare ground states away from criticality. It 
is also interesting to see in~\cref{fig:tfim_phase_transition_prediction} that 
the extremum of the VQE energy derivative and magnetisation derivative start 
out in a value above the point of the phase transition and then due to finite 
size effects converge with increasing $p$ to a value below it. This could 
imply that if only small system sizes are feasible low-depth VQE might locate 
the phase transition more accurately, because the light-cones of local 
observables don't span the whole system and hence finite size effects are not 
as strong.


On real hardware the errors present on NISQ quantum hardware and the sampling
noise lead to the barren plateau
~\cite{McClean_2018,Holmes_2021} phenomenon which will make optimising the VQE
cost functions harder. But our specific choice of ansatz circuits 
was found to exhibit only mild barren plateaus in the case of the 1D XXZ 
and 1D TFIM chains or the Heisenberg model on the kagome 
lattice~\cite{Wiersema_2020,Bosse_2022}. Furthermore, the fact that we run 
VQE for different circuit depths $p$ and at different points in the phase
diagram will help to warm start the VQE optimisation and use initial 
parameters that are already close to the optimal parameters and hence away 
from the barren plateaus~\cite{SmoothParameters_2022,Self_2021}. One could even 
consider simulating VQE on very small systems or with low circuit depths on  
classical computers and use the optimal parameters found there to warm start 
the optimisation on the real hardware.

Another open direction is to understand our results analytically, as far as 
possible. Dreyer, Bejan and Granet~\cite{Dreyer_2021} derive analytical 
scaling relations for local observables like $m_z$ and $m_x$ in 
the finite depth VQE states of the TFIM and show that scaling collapse 
techniques, similar to the bond dimension scaling in~\cite{Vanhecke_2019},
can be applied to order parameters measured in finite depth VQE states to
extract critical exponents. However, it is still open whether these techniques
also work for non-integrable models.
In~\cref{sec:vqe_hardness_in_noninteracting_free_fermion_systems} we also show
that in the case of translation-invariant free fermion systems deep circuits
are needed to connect ground states in different phases and that the circuit
depth required to prepare the ground state from a trivial initial state depends
on the spectral gap of the Hamiltonian; The smaller the gap is the deeper
circuits are required to prepare the ground state with high precision. Again,
it is open whether these findings also apply to non-integrable models.
The results of Chen, Gu and Wen~\cite{Chen_2010} give a partial answer 
in that short-time evolution with local unitaries can only connect ground states 
in the same phase. However, they allow for a bigger family of circuits than we
do and make no quantitative statements about how the circuit depth depends on
the gap of the target Hamiltonian.

\begin{acknowledgments}
  The authors would like to thank Filippo Gambetta and other
  members of the Phasecraft team for helpful discussions and feedback on the
  early drafts of this paper.
  This project has received funding from the European Research Council (ERC)
  under the European Union's Horizon 2020 research and innovation programme
  (grant agreement No.\ 817581) and from the EPSRC grant EP/S516090/1. 
  All data is  available at the University of Bristol data repository~\cite{Bosse2023_data}.
\end{acknowledgments}

\bibliography{paper}

\appendix

\section{Additional data} 
\label{sec:more_data}

\subsection{Additional data for the TFIM}
\label{sec:more_tfim_data}

\begin{figure}
  \begin{center}
    \input{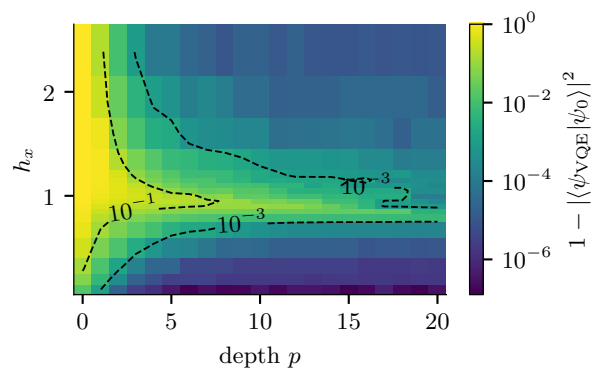}
  \end{center}
  \caption{Infidelity of the VQE state with the ground state as a function 
    of $p$ and $h_x$ for the 1D TFIM with chain length $L = 16$. 
  }
  \label{fig:tfim_vqe_fidelities}
\end{figure}

\begin{figure}
  \begin{center}
    \input{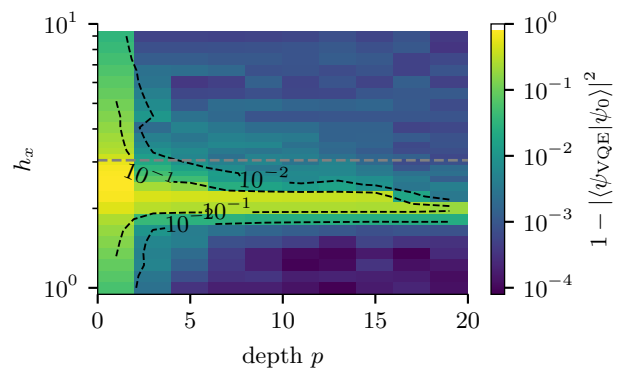}
  \end{center}
  \caption{Infidelity of the VQE state with the ground state as a function 
    of $p$ and $h_x$ for the 2D TFIM on $5 \times 5$ sites. 
  }
  \label{fig:2dtfim_vqe_fidelities}
\end{figure}

\begin{figure}
  \begin{center}
    \input{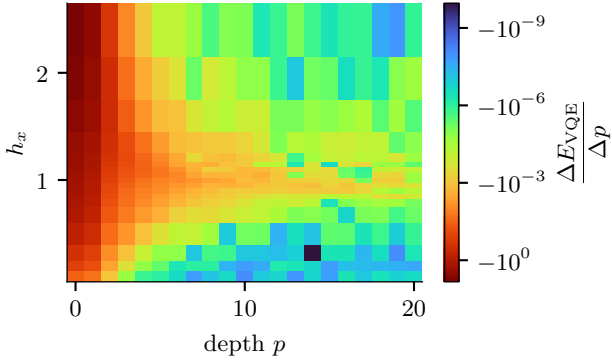}
  \end{center}
  \caption{VQE energy derivative as a function 
    of $p$ and $h_x$ for the 1D TFIM with chain length $L = 16$. 
  }
  \label{fig:tfim_energy_derivatives}
\end{figure}

\begin{figure}
  \begin{center}
    \input{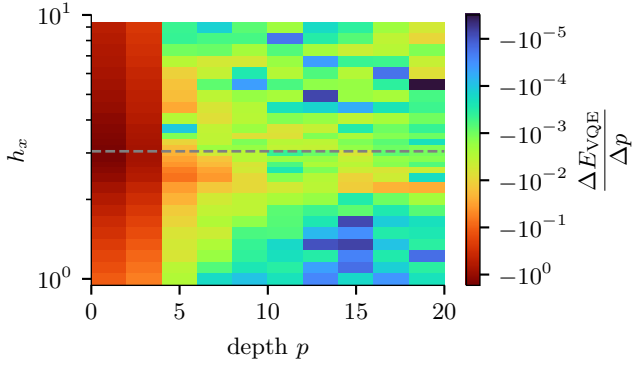}
  \end{center}
  \caption{VQE energy derivative as a function 
    of $p$ and $h_x$ for the 2D TFIM on $5 \times 5$ sites. 
  }
  \label{fig:2dtfim_energy_derivatives}
\end{figure}

In \cref{fig:tfim_vqe_fidelities,fig:2dtfim_vqe_fidelities} we show how the
fidelity of the VQE state 
with the exact ground state increases rapidly as we increase the ansatz depth $p$ 
far away from the phase transition, but increases much more slowly close to the 
phase transition. The same message can also be seen in 
\cref{fig:tfim_energy_derivatives,fig:2dtfim_energy_derivatives}
where the energy decreases rapidly for low $p$ far away from the phase transition, 
in particular for large $h_x$, but then converges soon for $h_x$ away from 1 
while convergence near the phase transition is slower, resulting in a ridge near 
$h_x = 1$ for the derivative of the VQE energy. Additionally, we can see 
how for large $p$ the energy derivative landscape becomes less smooth, due to 
the inability to resolve very small differences in the VQE energy.

\subsection{Additional data for the bilinear-biquadratic chain}
\label{sec:more_bilbiq_data}

\begin{figure}
  \begin{center}
    \input{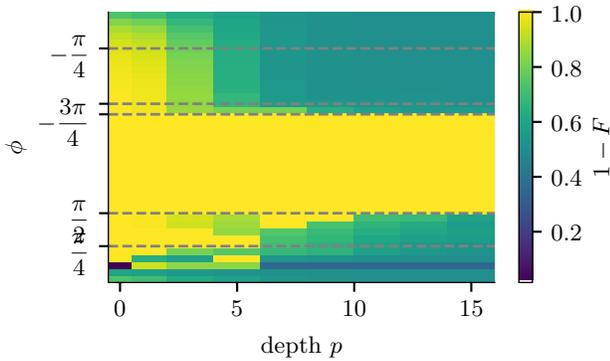}
  \end{center}
  \caption{Infidelity of the VQE state with the ground space as a function 
    of $p$ and $\phi$ for the bilinear-biquadratic chain on 12 sites. 
    Again, $F = \sum_{\textnormal{ground states} \ket{\psi_0}} |\braket{\psi_0 | \psi_{VQE}}|^2$ 
    denotes the squared overlap with the whole, possibly highly degenerate ground space 
    here. Note that compared to \cref{fig:tfim_vqe_fidelities} the colour 
    scale is on a linear scale here and not on a log scale, because the 
    infidelities are much larger for the 2D SSH model.
  }
  \label{fig:bilbiq_vqe_fidelities}
\end{figure}

\begin{figure}
  \begin{center}
    \input{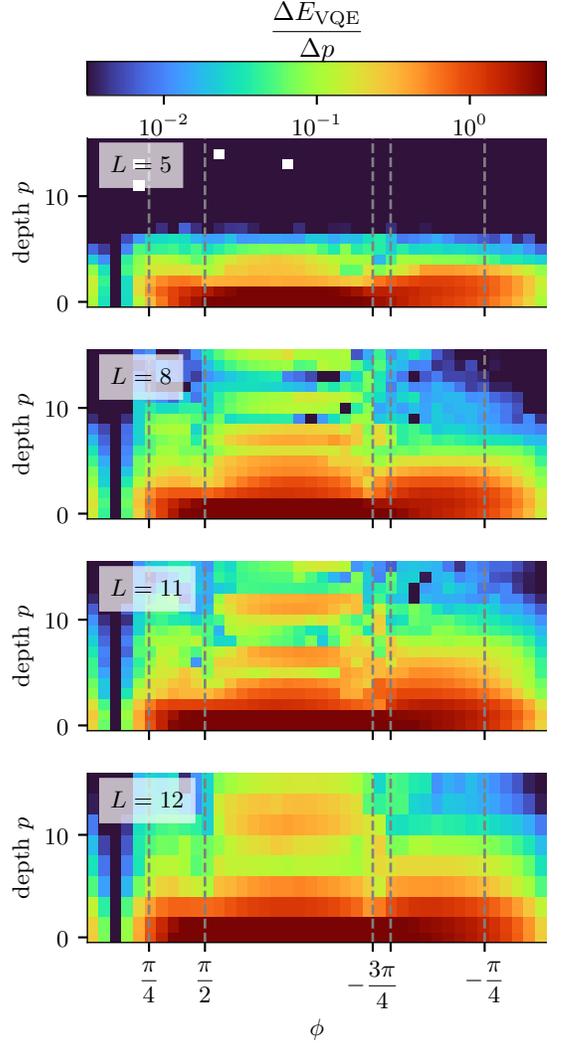}
  \end{center}
  \caption{VQE energy derivative for different circuit depths $p$ and
    different system sizes $L$ for the bilinear-biquadratic chain. For $L=5$
    the VQE energy converged for $p \geq 7$ for all $\phi$ and any non-zero
    value of $\Delta \EVQE / \Delta p$ is only due to numerical
    noise, hence we decided to cap the colour scale at $10^{-3}$. }
  \label{fig:bilbiq_vqe_hardness}
\end{figure}

In \cref{fig:bilbiq_vqe_fidelities} we show infidelity between the VQE state 
and the ground space for different ansatz 
depths $p$ and parameters $\phi$. We can clearly see, that except at the AKLT 
point where the initial VQE state is already the ground state the overlap between 
the VQE state and the exact ground space never converges to 1, irrespective of 
how large we allow $p$ to be. This makes it surprising that the different 
phases are so clearly visible in \cref{fig:bilbiq_vqe_hardness} where we compare 
the VQE energy derivatives of different system sizes. As expected, the VQE 
energy converges faster for smaller systems. But the qualitative features 
and signals between different system sizes are the same, although they are better
visible for larger system sizes.

\subsection{Additional data for the 2D SSH model}
\label{sec:more_ssh_data}

\begin{figure}
  \begin{center}
    \input{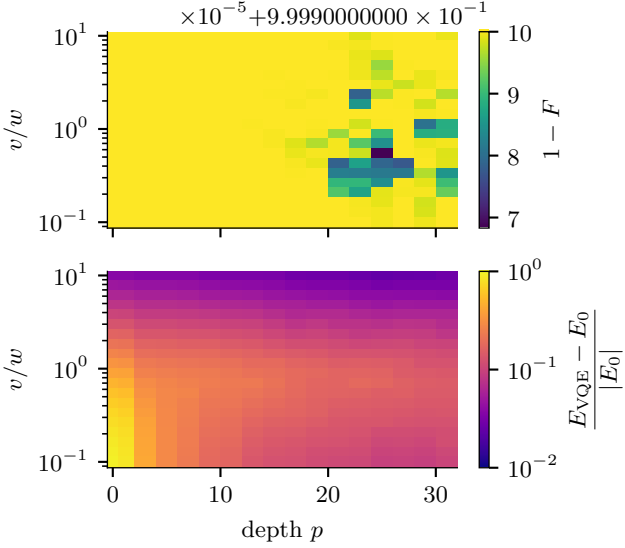}
  \end{center}
  \caption{(top pane) Infidelity of the VQE state with the ground space as a function 
    of $p$ and $v/w$ for the $10 \times 10$ SSH Model. 
    $F = \sum_{\textnormal{ground states} \ket{\psi_0}} |\braket{\psi_0 | \psi_{\textnormal{VQE}}}|^2$ 
    denotes the squared overlap with the whole, possibly highly degenerate ground space 
    here. Note that compared to \cref{fig:tfim_vqe_fidelities} the colour 
    scale is on a linear scale here and not on a log scale, because the 
    infidelities are much larger for the 2D SSH model.
    (bottom pane) VQE energy error as a function of $p$ and $v/w$ for the 
    $10 \times 10$ SSH Model. The energy error decreases as $p$ increases, 
    although the fidelity barely improves with $p$. Note also, that 
    the VQE energy error has a maximum as a function of $v/w$ for $p > 10$ 
    near the critical point $v/w$. However, noticing this requires a priori 
    knowledge of $E_0$ which is precisely the quantity that one hopes to find 
    using VQE.
  }
  \label{fig:ssh_vqe_fidelities}
\end{figure}

\begin{figure}
  \begin{center}
    \input{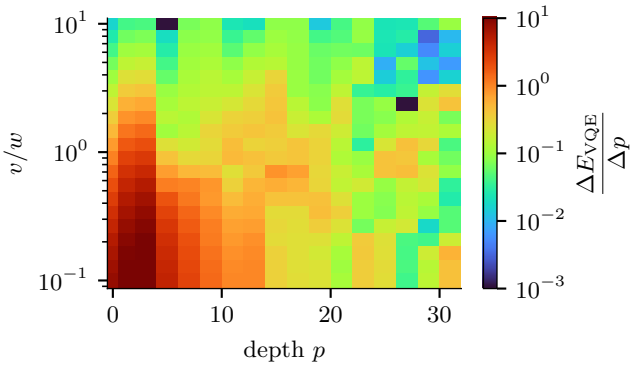}
  \end{center}
  \caption{VQE energy derivative as a function of circuit depth $p$ and $v/w$ 
    for the $10 \times 10$ 2D SSH model. The non-smooth values for $p \geq 20$
    with multiple minima are most likely due to the inability to find the global 
    ground state for all $p$ and $v/w$ and not due to the complex 
      phase diagram of the 2D SSH model.
  }
  \label{fig:ssh_energy_derivatives}
\end{figure}

In \cref{fig:ssh_vqe_fidelities} we show the infidelity between the VQE state 
and the exact ground space for all $p$ and $v/w$ for the $10 \times 10$ 2D SSH 
model. For most data points it is exactly 1 and only for a few points at 
$p \geq 20$ do we get a fidelity of $10^{-5}$. For our simulations of the $6 \times 6$ 
2D SSH model the fidelities were not quite that bad, for $p \geq 15$ we 
obtained fidelities of $0.2$ for $\frac{v}{w}$ away from 1. This is in contrast with 
\cref{fig:ssh_energy_derivatives} where we see a clear convergence of the energy 
with large improvements in the VQE energy, particularly in the topological phase 
at low $p$ and much lower derivatives at larger $p$. It should be noted that 
such very low fidelities are generically expected for large systems due to 
Anderson's locality ctatstrophe. In these cases the fidelity is a too strong 
measure of similarity between quantum states, since it bounds the error of 
\emph{all} possible observables, and not only that of local observables like 
we considered throughout this paper.

\section{The effect of parameter extrapolation} 
\label{sec:the_effect_of_parameter_extrapolation}

\begin{figure}
  \begin{center}
    \input{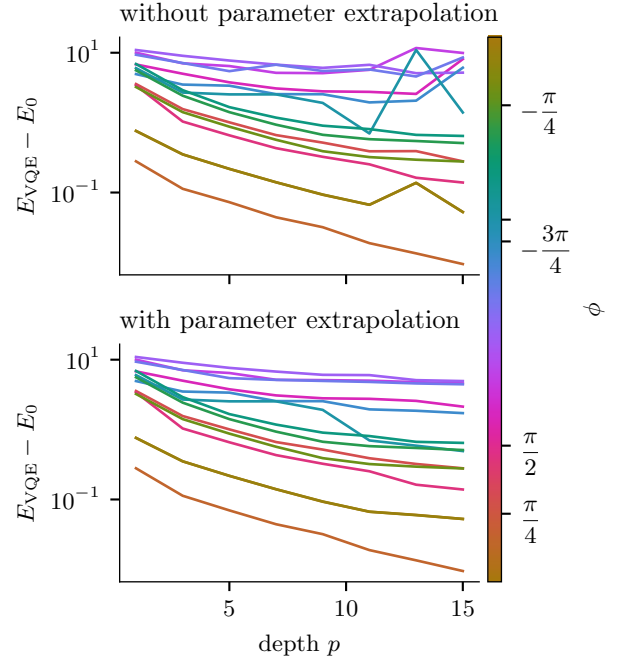}
  \end{center}
  \caption{
    The effect of parameter extrapolation in the bilinear-biquadratic chain.
    In the top subplot we show the final VQE energy errors of the best run 
    as a function of VQE depth $p$
    when using initial parameters randomly distributed in $[0, 1/p]$. In the 
    bottom subplot we show the VQE energy errors when using the parameter 
    extrapolation scheme described in \cref{sec:warm_starting_optimisation_with_previous_results}.
    Each line corresponds to a different value of $\phi$, here showing 13 runs 
    evenly spaced in $[0, 2\pi]$. The line closest to the AKLT point 
    $\phi = \arctan (1/3)$ is not shown here, because it is a constant at 
    $10^{-8}$ since here the initial state of VQE is already very close to the 
    target state.
  }
  \label{fig:bilbiq_energy_errors_iterative}
\end{figure}

\begin{figure}
  \begin{center}
    \input{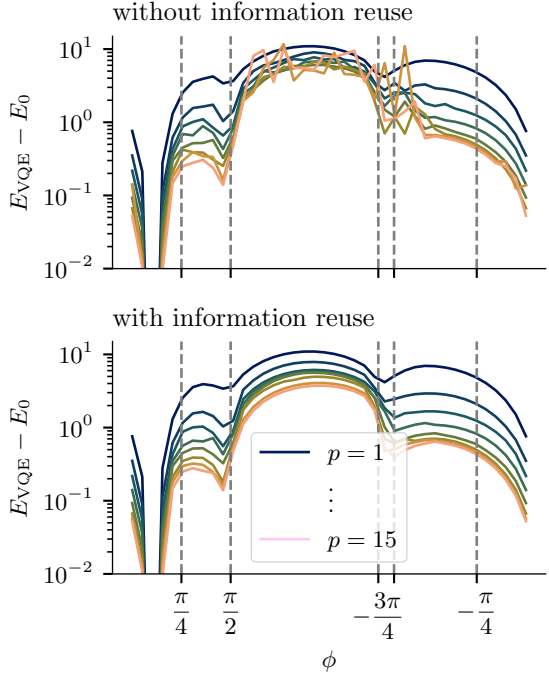}
  \end{center}
  \caption{
    The effect of information reuse in the bilinear-biquadratic chain. 
    In the top subplot we show the final VQE energy errors of the best run 
    when using initial parameters randomly distributed in $[0, 1/p]$, i.e. the 
    same data as in the top subplot of \cref{fig:bilbiq_energy_errors_iterative} 
    but with $\phi$ on the $x$-axis and each line corresponding to fixed $p$.
    In the bottom subplot we show the VQE energy errors when using the 
    information reuse scheme described in \cref{sec:warm_starting_optimisation_with_previous_results} 
    to warm start optimisation.
  }
  \label{fig:bilbiq_enregy_errors_infosharing}
\end{figure}

In \cref{fig:bilbiq_energy_errors_iterative,fig:bilbiq_enregy_errors_infosharing} 
we show the effect of our different schemes to warm start VQE with results from 
previous runs. \Cref{fig:bilbiq_energy_errors_iterative} shows the effect 
of extrapolating good low-depth parameters to higher depths where we got unlucky  
with the initial parameters. We can clearly see that with parameter extrapolation 
the energy decays monotonously as a function of $p$,
but not without. However, we found that for large $p$ 
the energy differences $\EVQE(p+1) - \EVQE(p+1)$
may get small enough that the inability to find the global minimum influences
the values significantly.

When reusing information at different $\phi$ to warm start VQE optimisation we 
find again, that this warm starting produces smoother curves than without,
although this time smooth as a function of $\phi$ (or $v/w$ for the 2D SSH model 
or $h_x$ for the TFIM) as exemplified in \cref{fig:bilbiq_enregy_errors_infosharing}. 
Again, the effect is more pronounced for larger $p$ where the search space 
is larger and the probability of finding suboptimal local minima is higher.

\section{Estimating phase transitions via fits to \texorpdfstring{$\EVQE(p)$}{EVQE}} 
\label{sec:estimating_phase_transitions_via_fits}

It was observed in previous work~\cite{Cade_2019,Biamonte_2020,Bosse_2022,Kattemoelle_2021,SmoothParameters_2022}
that often the ground state energy found by VQE decreases exponentially towards 
the true ground state energy as a function of the circuit depth $p$. 
Dreyer, Bejan and Granet~\cite{Dreyer_2021}, eq. (24) therein, also show this 
analytically for the TFIM when starting at $h_x = 0$ and remaining in the 
$h_x < 1$ phase. This motivates fitting a curve of the form 
\begin{equation}
  \EVQE(p) = a e^{-\gamma p} + E_{0,\textnormal{fit}}
  \label{eq:e_vqe_of_p}
\end{equation}
to the ground state energies found by VQE for a fixed Hamiltonian. $E_{0,\textnormal{fit}}$
is the predicted true ground state energy, $a$ the energy error of the initial 
state and $\gamma$ a measure for the hardness of preparing the true ground state.
Large $\gamma$ means that the VQE energy quickly approaches the true ground state
energy and shallow circuits suffice to prepare the true ground state whereas 
small $\gamma$ implies one needs deep circuits to prepare the true ground state. 
Since shallow circuits cannot prepare states with long-range entanglement,
but the states  at phase transitions have long-range entanglement one may 
conjecture that $\gamma$ is small precisely at the phase transitions. We found 
that this holds for the TFIM and bilinear-biquadratic chain and that fitting 
with \cref{eq:e_vqe_of_p} is another method that only needs the VQE energy at
different depths $p$ to detect phase transitions. However, for the 2D SSH model, 
we found this method signals the phase transition less clearly than the VQE
energy derivative.

\section{VQE hardness in non-interacting fermionic systems with translation invariance}
\label{sec:vqe_hardness_in_noninteracting_free_fermion_systems}

In this section we show that if $\ket{\psi_i}$ is the many-particle ground state
of a translation invariant, local, non-interacting fermionic Hamiltonian $H^{(1)}$ and we 
wish to prepare  from this the ground state of another such Hamiltonian $H^{(2)}$ 
via evolution with said Hamiltonians this is possible in constant time 
iff the Bloch bundles defined by the ground states of $H^{(1)}$ and $H^{(2)}$
are isomorphic. Furthermore we show that if $H^{(1)}$ is trivial in the sense 
that its eigenstates are simple Fourier modes the time needed to 
prepare the ground state of $H^{(2)}$ depends on the band gap $\Delta E$ of $H^{(2)}$.

Let's consider two non-interacting fermionic system in $d \leq 3$ spatial dimensions 
and $M \in \mathbb{N}$ sites per unit cell (or $d \geq 4$ spatial dimensions
and $M=1$ site per unit cell), described by single-particle, translation-invariant Hamiltonians
\begin{equation}
  H^{(i)} = \sum_{\vec{x}, \vec{y}} \sum_{m,n} 
  h^{(i)mn}_{\vec{x} - \vec{y}} \ket{\vec{x},m}\bra{\vec{y},n}.
\end{equation}
where $\vec{x}$ and $\vec{y}$ are the unit cells of a square lattice in $d$ 
dimensions and $m$ and $n$ label the $M$ sites in each unit cell.
Using the Fourier transform
\begin{equation}
  \ket{\vec{k}} = \sum_{\vec{x}} e^{i\vec{k}\vec{x}} \ket{\vec{x}}
\end{equation}
they can be block-diagonalised into 
\begin{equation}
  H^{(i)} = \int \dk \sum_{mn} H^{(i)mn}(\vec{k}) \ket{m}\bra{n} \otimes \ket{\vec{k}}\bra{\vec{k}}
\end{equation}
where the integral goes over the first Brillouin zone $\mathbb{T}^d$ and  
\begin{equation}
  H^{(i)mn}(\vec{k}) = \sum_{\Delta \vec{x}} e^{-i\vec{k} \Delta \vec{x}} h^{(i)mn}_{\Delta \vec{x}}.
\end{equation}
Now each block $H^{(i)}(\vec{k})$ can be diagonalised via some $U^{(i)}(\vec{k})$ with 
eigenvalues $E_\mu^{(i)}(\vec{k})$ s.t. 
\begin{equation}
  \ket{\mu^{(i)}(\vec{k})} = \sum_m U_{\mu m}^{(i)}(\vec{k}) \ket{m}
\end{equation}
and thus
\begin{equation}
  H^{(i)} =  \int \dk \sum_{\mu} E_{\mu}^{(i)}(\vec{k}) 
    \ket{\vec{k}, \mu^{(i)}(\vec{k})}\bra{\vec{k}, \mu^{(i)}(\vec{k})}.
\end{equation}

These single-particle eigenstates $\ket{\vec{k}, \mu^{(i)}(\vec{k})}$ can be used to construct 
the corresponding \emph{Bloch bundles} $E^{(i)}$ as sub-bundles of the trivial bundle 
$\mathbb{T}^d \times \mathbb{C}^M$ with projection 
$\pi_{1} : \mathbb{T}^d \times \mathbb{C}^M \to \mathbb{T}^d$.
The base space of the Bloch bundles $E^{(i)}$ is the first Brillouin zone 
$\mathbb{T}^d \ni \vec{k}$ and the fibers are
\begin{equation}
  \pi^{(i) -1}(\vec{k}) = \textrm{span}\left\{ \ket{\vec{k}, \mu^{(i)}(\vec{k})} \middle| \mu=1,\cdots,N_e \right\}
\end{equation}
where $1 \leq N_e \leq M$ is the number of filled bands and dimension of the vector bundle 
and $\pi^{(i)}: E^{(i)} \to \mathbb{T}^d$ are the bundle projection maps. If 
$h^{(i)mn}_{\vec{x}-\vec{y}}$ decays rapidly as a function of $\vec{x}-\vec{y}$, then $H^{(i)}(\vec{k})^{mn}$ 
will be smooth as a function of $\vec{k}$. If, furthermore, $H^{(i)}(\vec{k})$ has a gap 
between the $N_e$-th and the $(N_e + 1)$-th eigenvalue, then the fibers 
$\pi^{(i) -1}(\vec{k})$ (interpreted as subspaces of $\mathbb{C}^M$) depend smoothly on $\vec{k}$ and 
the $E^{(i)}$ are indeed smooth sub-bundles of the trivial bundle.

A translation invariant circuit that takes the ground state of $H^{(1)}$ to the 
ground state of $H^{(2)}$ is now given by a family of unitaries parameterised by 
$\vec{k}$ that takes the fiber at $\vec{k}$ in $E^{(1)}$ to the corresponding fiber in 
$E^{(2)}$, i.e. a $W(\vec{k})$ s.t. 
\begin{equation}
  W(\vec{k}) \pi^{(1)\, -1}(\vec{k}) = \pi^{(2)\, -1}(\vec{k}).
  \label{eq:W_of_k}
\end{equation}
One possible choice for $W(\vec{k})$ is 
$W(\vec{k}) = U^{(2)}(\vec{k}) U^{(1)\,-1}(\vec{k})$. However since we only
need~\cref{eq:W_of_k} and not $W(\vec{k}) \ket{\mu^{(1)}(\vec{k})} = \ket{\mu^{(2)}(\vec{k})}$
a different choice might yield a smoother $W(\vec{k})$.
If $W(\vec{k})$ smoothly depends on $\vec{k}$~\cref{eq:W_of_k} is equivalent to
saying that the induced map $W|_{E^{(1)}}: E^{(1)} \to E^{(2)}$ is a bundle
isomorphism. In $d \leq 3$ or $M=1$ such a bundle isomorphism exists if and
only if the Chern numbers of $E^{(1)}$ and $E^{(2)}$ are equal~\cite{Panati_2007}.

We can also use this $W(\vec{k})$ to construct the effective Hamiltonian whose 
time evolution takes the groundstate of $H^{(1)}$ to that of $H^{(2)}$: 
In the $(\vec{k},m)$ basis define it via 
\begin{equation}
  \Gamma(\vec{k}) = i \log W(\vec{k}) \Leftrightarrow W(\vec{k}) = e^{-i \Gamma(\vec{k})}.
\end{equation}
Since basis changing (here the inverse Fourier transformation) and taking 
logarithms commutes we can directly write down $\Gamma$ in the $(\vec{x},m)$ basis as 
\begin{equation}
\begin{aligned}
  \Gamma &= \sum_{\vec{x},\vec{y},m,n} \Gamma^{mn}_{\vec{x}-\vec{y}} \ket{\vec{x},m}\bra{\vec{y},n} \\ 
         &= \sum_{\vec{x},\vec{y},m,n} \left(\int \dk e^{ik(\vec{x}-\vec{y})} \Gamma(\vec{k})^{mn} \right) \ket{\vec{x},m}\bra{\vec{y},n}\\
         &= \sum_{\vec{x},\vec{y},m,n} \left( \int \dk e^{ik(\vec{x}-\vec{y})} \left(i \log W(\vec{k})\right)^{mn} \right) \ket{\vec{x},m}\bra{\vec{y},n}.  
\end{aligned}
\end{equation}
If $W(\vec{k})$ is smooth as a function of $\vec{k}$, then $\log W(\vec{k})$ is
also smooth and by the Paley-Wiener theorem 
$\Gamma_{\vec{x}-\vec{y}}^{mn} = \int \dk e^{i\vec{k}(\vec{x}-\vec{y})} \left(i \log W(\vec{k})\right)^{mn}$
decays exponentially as a function of $\vec{x}-\vec{y}$. This implies that $\Gamma$ is local 
in real space and hence by a Trotter + Lieb-Robinson argument also found 
in \cite{Chen_2010} (approximate) time evolution with it can be implemented in
constant time.

The circuits produced by the Hamiltonian Variational ansatz are also generated 
by the evolution with translation-invariant, non-interacting fermionic
Hamiltonians. Hence, if no such $W(\vec{k})$ generated  by the evolution with 
a short-ranged effective Hamiltonian $\Gamma$ exists, it is not possible to prepare the ground state of $H^{(2)}$ from $\ket{\psi_i}$ with low 
circuit depth $p$ using VQE with the HV ansatz.

As we have discussed, the existence of a short range effective Hamiltonian
$\Gamma$ depends entirely on the analytic properties of $W(\vec{k})$. The
analyticity of $W(\vec{k})$ is related with the existence of an energy gap in
the path connecting $H^{(1)}$ and $H^{(2)}$. This can be shown to be the case
if $H^{(1)}$ is trivial in the sense that its eigenstates are simple
Fourier modes (i.e. $U^{(1)}(k) = \mathrm{id}$). Then the Wannier functions of
$H^{(2)}$ are given by 
\begin{equation}
  \ket{w_{\vec{x},\mu}} = \sum_{\vec{y},m} \int \dk \, e^{i\vec{k}(\vec{x}-\vec{y})} W(\vec{k})^{\mu m} \ket{\vec{y}, m}.
\end{equation}
Theorem 2.6 in \cite{Hastings_2006} then implies that these Wannier functions 
decay as $\braket{\vec{y},m | w_{\vec{x},\mu}} \leq C \cdot e^{-\gamma |\vec{x}-\vec{y}|}$ where 
$\gamma$ grows with the band gap $\Delta E$ of $H^{(2)}$. This implies that 
in this case also the effective Hamiltonian $\Gamma$ is more strongly localized 
if $H^{(2)}$ has a large band gap. By the same Trotter + Lieb-Robinson arguments 
as above this also implies that deep circuits are required 
to prepare the groundstate of $H^{(2)}$ that have a small gap if we start 
with a trivial Slater determinant of Fourier modes $\ket{\psi_i}$.

\end{document}